\documentclass{emulateapj}
\shorttitle{Clusters in the Galactic Anticenter Stellar Structure}
\shortauthors{Frinchaboy et al.}
\begin{document}

\title{Star Clusters in the Galactic Anticenter Stellar Structure and the
  Origin of Outer Old Open Clusters}

\author{Peter M. Frinchaboy\altaffilmark{1}, Steven R.
  Majewski\altaffilmark{1}, Jeffrey D. Crane\altaffilmark{1},
  I. Neill Reid\altaffilmark{2}, Helio J. Rocha-Pinto\altaffilmark{1},
  Randy L. Phelps\altaffilmark{3}, Richard J. Patterson\altaffilmark{1}
  and Ricardo R. Mu\~{n}oz\altaffilmark{1}}

\email{pmf8b, srm4n, jdc2k, hjr8q, rrm8f, rjp0i@virginia.edu, \\
  inr@stsci.edu, phelps@csus.edu}

\altaffiltext{1}{Astronomy Dept., Univ.\ of Virginia, Charlottesville,
  VA 22903}

\altaffiltext{2}{Space Telescope Science Institute, Baltimore, MD 21218}

\altaffiltext{3}{Physics \& Astronomy Dept., CSUS, Sacramento, CA 95819}

\begin{abstract}
  
  The Galactic Anticenter Stellar Structure (GASS) has been identified
  with excess surface densities of field stars in several large area sky
  surveys, and with an unusual, string-like grouping of five globular
  clusters.  At least two of these are diffuse, young ``transitional''
  clusters between open and globular types.  Here we call attention to
  the fact that four {\it younger} open or transitional clusters extend
  the previously identified, string-like cluster grouping, with at least
  one having a radial velocity consistent with the previously found GASS
  velocity-longitude trend.  All nine clusters lie close to a plane
  tipped 17$\arcdeg$ to the Galactic plane.  This planar orientation is
  used to forage for additional potential cluster members in the inner
  Galaxy, and a number are found along the same plane and string-like
  sequence, {\it including almost all fifteen known outer, old open clusters}.
  Tidal accretion of a dwarf satellite
  galaxy on a low inclination orbit --- perhaps the GASS system ---
  appears to be a plausible explanation for the origin of the outer, 
  old open and transitional clusters of the Milky Way.  We use these
  clusters to explore the age-metallicity relation of the putative 
  accreted GASS progenitor.  Finally, we provide the first radial
  velocity of a star in the cluster BH 176 and discuss its implications.

\end{abstract}
\keywords{Galaxy: structure---Galaxy: disk---galaxies: interactions---
Galaxy: open clusters and associations: general----Galaxy: globular clusters: general}

\section{Introduction}

Excesses of stars beyond the apparent limit of the Galactic disk have
been used to argue for the presence of a distinct, extended stellar
structure wrapping around the disk at low latitudes (Newberg et al. 2002;
Ibata et al. 2003, hereafter I03; Majewski et al., hereafter M03;
Yanny et al. 2003, hereafter Y03; Rocha-Pinto et al., hereafter R03).
Collectively,
these surveys suggest that the structure spans $|b|<30\arcdeg$ and
at least $122\arcdeg<l<225\arcdeg$ 
at a mean
$R_{GC}\sim16$ kpc (I03, R03) and radial thickness $\lesssim4$ kpc (Y03,
I03).  However, because of unfortunate placement behind considerable
extinction, it has been difficult to get information on the system's
true shape, orientation, breadth, etc.  Even the location of the
structure's center (presumably corresponding to a ``nucleus'') remains
uncertain; thus we refer to the entire system here as the Galactic
anticenter stellar structure (GASS).

The origin of GASS --- originally described as a ``ring'' around the
Galaxy (I03, Y03) --- is also not definitively established, with a
number of potential scenarios outlined, e.g., by I03: a tidally
disrupted satellite galaxy, an outer spiral arm, or (their
preference) a resonance induced
by an asymmetric Galactic component.  From among possibilities involving
accretion, \citet{helmi} explore the extremes of dynamically young and
old tidal debris, and \citet[ hereafter C03]{crane} argue, as did Y03 earlier, that the most
straightforward interpretation is that GASS is a
disrupted satellite galaxy, resembling in many ways the 
Sagittarius (Sgr) dwarf galaxy (e.g., M03) system.
As evidence,
C03 point to: (1) a velocity-longitude trend indicating
a slightly non-circular orbit, (2) a velocity dispersion smaller than
even that of thin disk stars, (3) a wide metallicity spread from [Fe/H] $=-1.6
\pm 0.3$ (Y03) to at least 
$=-0.4 \pm 0.3$ dex, and (4) at least four
star clusters apparently associated with the stream based on 
position {\it and} radial velocity (RV).  These clusters (Pal 1, NGC
2808, NGC 5286, NGC 2298), plus a fifth having no RV measurement (BH
176), lie in an unusual, arc-like configuration not seen elsewhere among
low latitude, outer globular clusters (GCs), but one resembling
configurations expected for tidal debris systems
\citep[e.g.,][]{bellazzini}.

Several unusual GCs are identified by C03 as potential
GASS members.
Pal 1 is both very small ($M_v =
-2.5$) and relatively metal-rich ([Fe/H] $= -0.6$) for an $R_{GC} > 8$
kpc GC.  \citet{rosenberg} derive a Pal 1 age of $8\pm2$ Gyr and
suggest it is either the youngest GC or one of the oldest open
clusters (OCs) in the Galaxy.  \citet[ hereafter PS03]{phelps_schick} find BH
176 
to be young ($7.0 \pm 1.5$ Gyr) and metal-rich ($-0.20 \le$
[Fe/H] $\le +0.20$) and suggest it is ``transitional'' between a young,
metal-rich GC and a massive, metal-rich old OC.

Prompted by this hint that GASS may contain younger, more metal-rich 
star clusters, and by our new distance for OC Saurer A (Sau A)
\citep{fp02} placing it near the R03 and C03 tracings of GASS, we
search here for other {\it open clusters} coincident with GASS,
and find an interesting potential connection 
with the Milky Way (MW) old OC system.

\section{Radial Velocity of a BH 176 Giant Candidate }

Because BH 176 has no measured RV to check
against the apparent $l-v_{gsr}$ trend of GASS, observations of
candidate BH 176 giant stars were obtained with the R-C Spectrograph and
600 line mm$^{-1}$ grating (4.3 \AA\ per resolution element) on the 
CTIO 1.5-m telescope on UT 2003 August
02.  Three obvious stars 
along the BH 176 red giant sequence, redder than the typical field stars,
and at the cluster center
were selected from the Ortolani et al. (1995, OBB hereafter) 
database.  However, only the spectrum of the star at $(\alpha,
\delta)_{J2000.0} = $(15:39:07.8,$-$50:03:11) proved of sufficient quality
for a reliable RV.  
The spectrum from $4400-5240$ \AA\ was cross-correlated against
spectra of stars Gl 803 (spectral type M0) and Gl 643 (M3.5), using both
IRAF's {\sc fxcor} task and our own software (C03).  Weak MgH+Mgb
absorption in the spectrum of the target star strongly suggests that it
is a giant (of spectral type M2-M3).  An average $v_{hel} = 85$ km
s$^{-1}$ is obtained, where a 30 km s$^{-1}$ error is estimated from the
spread of results using different RV standards and software.  While not
highly precise, and of only one giant star in the cluster field, the
$v_{gsr} = -27\pm30$ for this star tantalizingly suggests (if
it is a BH 176 member) that BH176 follows the $l-v_{gsr}$ trend
of GASS (C03; see Fig.\ 2 below).  Therefore, we include BH176 among
more likely GASS clusters.

\begin{figure}
\epsscale{0.90}
\plotone{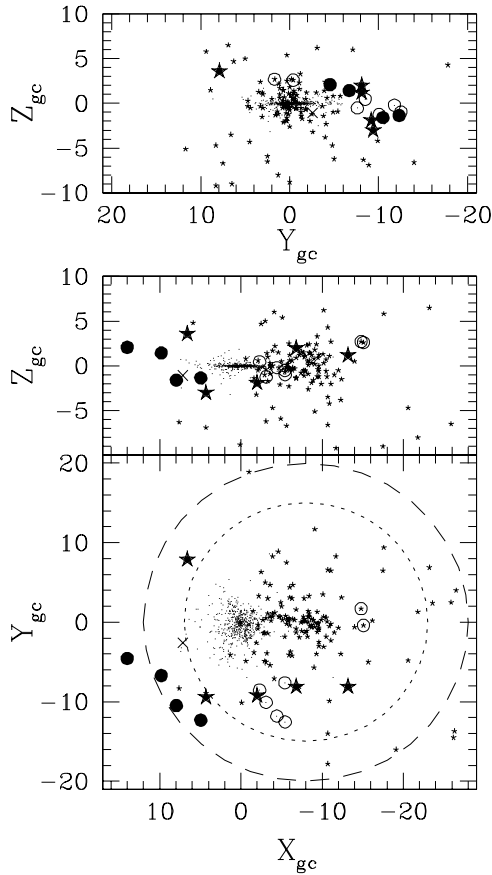}
\caption{Cartesian distribution (in kpc) of clusters in the heliocentric,
  left-handed Galactic system (see M03).  Dots show open clusters from
  \citet{dias}, stars the globular clusters from \citet{harris}.  Large
  stars are the five GASS clusters from C03.  Large filled circles are
  the four new GASS cluster candidates discussed in \S3.  Open circles
  show seven GASS clusters selected by plane-fitting (\S4).  Cross
  denotes Berkeley 22 (see \S5). }
\end{figure}

\section{Distant Open Clusters Correlated to GASS}

Fig.\ 1 shows the
distribution of OCs \citep[using the on-line database by]
[ updated as in Table 1]{dias} and GCs \citep[from the 
latest on-line compilation by][]{harris}.
The five GASS cluster candidates from C03 are marked by large star
symbols.\footnote{The globulars NGC 1851 and NGC 1904 lie along the same
  $X_{GC}-Y_{GC}$ cluster trend and are coincident with the R03 tracing
  of GASS, but C03 show them to have RVs and/or $Z_{GC}$ discordant with
  those of the GASS M giants and five other GCs.  
}  The four OCs with $R_{GC} > 15$ kpc --- AM-2, Tombaugh 2, Berkeley 29 and
Sau A --- are {\it also} exceptional for lying in a string-like
configuration (large circles in Fig.\ 1).  The unusual $R_{GC}$ of the
first three have long been recognized
\citep[e.g.,][OBB]{adler,kaluzny}, but the extreme $R_{GC}$ of
Sau A has only recently been noted \citep{fp02}.  Considering spatial
biases in the known OC sample (due, e.g., to extinction), four
clusters in one part of the sky might not be considered too unusual.
However, these four extreme OCs also lie along the
GASS M giants (see Fig.\ 4 of R03 and Fig.\ 1 of C03) and extend the
spatial trend of the GASS GCs from C03.  Moreover, 
Tombaugh 2's measured RV, $v_{gsr} = -74.8$
km s$^{-1}$, places it squarely on the $l-v_{GSR}$ trend observed for
GASS stars and clusters 
(Fig.\ 2).  In addition, Sau A
\citep{fp02,carraro} and AM-2 \citep[OBB]{lee} have ages and
metallicities (Table 1) similar to Pal 1.  AM-2, like BH 176 and Pal 1,
has been discussed as a possible ``transitional cluster'' (TC) by OBB.
These similarities strongly suggest these four ``open"
clusters may also be associated with GASS.

These nine ``primary'' GASS candidate clusters --- though spread over
dozens of kiloparsecs, three Galactic quadrants, and $-3.0 < Z_{GC} < +3.6$
kpc (Table 1 gives adopted coordinates) 
--- also lie close to a single, inclined plane: a least squares
fit finds all nine within 2.35 kpc of $0.057 X_{GC} - 0.297
Y_{GC} + 0.953 Z_{GC} = 2.521$, with an RMS of only 1.39 kpc.
Because these nine clusters were
first identified on the basis of their $X_{GC}-Y_{GC}$ configuration
(though, admittedly, clusters with extreme $Z_{GC}$ were ruled
out by C03), 
and further winnowing of the sample relied only on available
RVs (C03), there should be no reason to expect these
clusters to lie as close to one plane as they do; yet 
this RMS
is smaller than that for association to the {\it Galactic} plane (GP):
2.15 kpc.  Cohesion to this plane (with pole $[l,b]=[79.2,
-72.7]{\arcdeg}$) is further support for dynamical association of these
clusters.

\begin{figure}
\epsscale{1.00}
\plotone{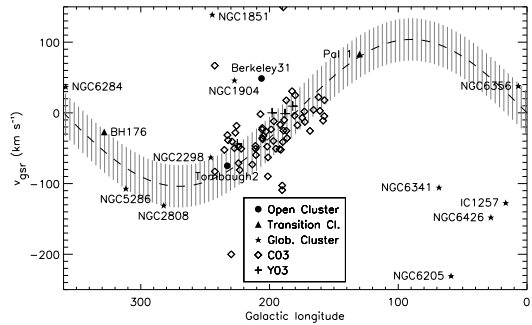}
\caption{The $l-v_{GSR}$ distribution of objects lying
  within 2.35 kpc of the GASS cluster plane and other clusters discussed
  in C03.  Corrections to $v_{gsr}$ assume a solar apex of
  $(\alpha,\delta)=(18h,30\arcdeg)$ at 20.0 km s$^{-1}$ and an LSR
  rotation of 220 km s$^{-1}$.  The hash marks represent a velocity
  dispersion of 30 km s$^{-1}$ about the $v_{gsr}$ of a circularly
  orbiting object at r$_{GC}=18$ kpc with $v_{circ}=220$ km s$^{-1}$,
  which approximately matches GASS M giant velocities.  
}
\end{figure}

\begin{deluxetable*}{lcrrrrrrrrrrrrr}
\tabletypesize{\scriptsize}
\tablecaption{Cluster Data}
\tablewidth{0pc}
\tablehead{ \colhead{Cluster} & \colhead{Type} & \colhead{$l$($\arcdeg$)}
& \colhead{$b$($\arcdeg$)}  & \colhead{$R_{gc}$} & \colhead{$d_{\sun}$}
& \colhead{$X_{GC} $} & \colhead{$Y_{GC} $}& \colhead{$Z_{GC} $}& \colhead{Age}
& \colhead{[Fe/H]} & \colhead{$v_{gsr}$}& \colhead{$M_V$}& \colhead{DP}& \colhead{Ref}}

\startdata
Pal 1        & TC & 130.07 &  19.03 &  17.00 &  10.90 &   6.60 &   7.90 &   3.60 &  8.00   & $-0.60$   & $-107.0$  & $-2.47$   & $-1.06$ & a \\
NGC 2298     & GC & 245.63 & $-16.01$ &  15.70 &  10.70 &   4.30 &  $-9.40$ &  $-3.00$ & 12.59   & $-1.85$   & $-62.8$   & $-6.30$   & $-2.35$ & b \\
NGC 2808     & GC & 282.19 & $-11.25$ &  11.10 &   9.60 &  $-2.00$ &  $-9.20$ &  $-1.90$ &  9.12   & $-1.15$   & $-130.8$  & $-9.39$   & $-1.72$ & b \\
NGC 5286     & GC & 311.61 &  10.57 &   8.40 &  11.00 &  $-7.20$ &  $-8.10$ &   2.00 & \nodata & $-1.67$   & $-107.0$  & $-8.61$   & $-1.06$ &   \\
BH 176       & TC & 328.41 &   4.34 &   9.70 &  15.60 & $-13.20$ &  $-8.10$ &   1.20 &  7.00   &  0.00   & $-27.0$   & $-4.35$   &  1.12 & c \\
\hline
Arp-Madore 2 & TC & 248.12 &  $-5.87$ &  17.91 &  13.34 &   4.94 & $-12.31$ &  $-1.36$ &  5.00   & $-0.50$   & \nodata & \nodata &  0.11 & d \\
Berkeley 29  & OC & 197.98 &   8.02 &  22.56 &  14.87 &  14.00 &  $-4.54$ &   2.07 &  1.06   & $-0.18$   & \nodata & $-4.64$   &  1.60 & e\\
Saurer A     & OC & 214.31 &   6.83 &  19.08 &  11.97 &   9.81 &  $-6.70$ &   1.42 &  7.20   & $-0.50$   & \nodata & \nodata &  1.38 & f \\
Tombaugh 2   & OC & 232.83 &  $-6.88$ &  19.15 &  13.26 &   7.95 & $-10.49$ &  $-1.58$ &  2.00   & $-0.36$   & $-74.0$   & \nodata & $-0.47$ &  \\
\hline
ESO 092-18   & OC & 287.12 &  $-6.65$ &  11.26 &  10.60 &  $-3.11$ & $-10.06$ &  $-1.22$ &  1.05   & \nodata & \nodata & \nodata & $-0.88$ &  \\
ESO 093-08   & TC & 293.50 &  $-4.04$ &  12.82 &  13.70 &  $-5.46$ & $-12.53$ &  $-0.96$ &  4.47   & $-0.40$   & \nodata & \nodata &  0.02 & c \\
Saurer C     & OC & 285.09 &   2.99 &  10.25 &   8.83 &  $-2.30$ &  $-8.51$ &   0.46 &  2.82   & \nodata & \nodata & \nodata &  0.31 &  \\
Shorlin 1    & OC & 290.56 &  $-0.92$ &  12.32 &  12.60 &  $-4.43$ & $-11.79$ &  $-0.20$ & \nodata & \nodata & \nodata & \nodata &  0.53 &  \\
BH 144       & OC & 305.34 &  $-3.15$ &   8.06 &   9.35 &  $-5.46$ &  $-7.61$ &  $-0.51$ &  0.67   & \nodata & \nodata & \nodata & $-1.06$ &  \\
NGC 6284     & GC & 358.35 &   9.94 &   7.60 &  15.30 & $-15.10$ &  $-0.40$ &   2.60 & \nodata & $-1.32$   & 37.2    & $-7.97$   & $-0.78$ &  \\
NGC 6356     & GC &   6.72 &  10.22 &   7.60 &  15.20 & $-14.80$ &   1.70 &   2.70 & \nodata & $-0.50$   & 38.1    & $-8.52$   & $-1.29$ &  
\enddata
\tablerefs{All distances (including ``DP'', the distance from the plane
defined in \S3) are in kpc, ages are in Gyr, and $v_{gsr}$ is in km s$^{-1}$.
All values are from \citet{harris} or \citet{dias} except as noted:
a. age from \citet{rosenberg};
b. age from \citet{sw02};
c. $d_{\sun}$, age, [Fe/H] from PS03;
d. age, [Fe/H] from \citet{lee};
e. $M_V$  from \citet{lata}.
f. [Fe/H] from \citet{carraro}.
}
\end{deluxetable*}

\section{Foraging for Other Potential Members}

The alignment of the nine primary GASS clusters can be extrapolated to
search for additional associated clusters in the more crowded inner
Galactic regions.  This exercise, while intended simply to identify
other interesting possible members for future study, does turn up an
interesting coincidence (\S5).  For now we exclude consideration of: (1)
the
many clusters with $d_{\sun}<7$ kpc, many of which, by chance, fall near
the GASS cluster plane because its line of nodes with the GP lies nearby,
and (2) the populous ``disk'' GC system with
$R_{GC} < 7$ kpc.  Adopting this conservative ``Volume of
Avoidance'' (VoA) is consistent with: (1) an expectation that, if tidal
debris, GASS must arc to the other side of the Galactic Center, and
(2) a presumption, based on the $X_{GC}-Y_{GC}$ distribution of the GASS
clusters and M giants, along with the C03 velocities, that the GASS orbit is only
slightly elliptical with perigalacticon $\gtrsim 7$ kpc.
A search through the cluster catalogues for other objects within 2.35
kpc of the best-fit plane (2.35 kpc is the largest deviation of the nine
plane-defining clusters) yields six more OCs and six more GCs.

The total sample of 21 clusters have an RMS of only 1.03 kpc about the \S3 plane,
and represent an overdensity of clusters along any plane.
Statistical tests of the parent sample and 
scrambled versions thereof
find randomly placed planes to typically have only 8.2 clusters within 2.35 kpc.  
The cluster catalogues were scrambled to preserve the combined density law 
and discovery function (and tested with the VoA above):
GCs were randomly rotated about the $Z$-axis 
through the Galactic Center, preserving $R_{GC}$ and $Z_{GC}$, 
while OCs were randomly rotated about the $Z_{GC}$-axis, 
preserving $Z_{GC}$ and $d_{\sun}$.
The tests show that probability of finding 20 or more clusters in a 
Poisson distribution with mean of only 8.2 is 0.04\%.
However, these statistics also suggest $8 \pm \sqrt{8}$ chance
``interlopers'' lie among the 21 clusters.  Seven of the new clusters have
RVs useful to prune (as in C03) to the most interesting candidates based
on correlation to the previously found $l-v_{GSR}$ trend (Fig.\ 2).
Thus, we ``demote'' as less likely to be associated the globulars NGC
6205, NGC 6341, NGC 6426 and IC 1257 and the OC 
Berkeley 31, but find the globulars NGC 6284 and NGC 6356 to nicely fall
along the Fig.\ 2 trend.  The latter two GCs, together with the
five new clusters without RVs plus nine primary GASS clusters (i.e., all
clusters in Table 1) collectively define an asymmetric
distribution in space: For example, 14 of the 16 clusters have
$Y_{GC}<0$ and almost all define an arcing sequence in various Fig.\ 1
projections, strengthening the impression of an inclined, tidal
tail-like trail in three-dimensional space (Fig.\ 1).  The seven new
candidate clusters are actually {\it even more} tightly confined to the
nominal plane (RMS $=0.81$ kpc, and with all clusters within 1.3 kpc) than
are the nine clusters that defined it (RMS $=1.39$ kpc)!  The net RMS about
the \S3 plane for all Table 1 clusters is 1.15 kpc.

Despite some affinity of the outer clusters for
the GP, only 16 from the parent population have
$Z_{GC}<2.35$ kpc, and these with a larger RMS($Z_{GC}$) $=1.31$ kpc.
Thirteen of these 16 ``GP clusters'' overlap with the sample of 21
above.  The probability of finding more than 21 clusters in a plane
from a parent population whose average is 16 is 13.2\%.
While it might still be argued that the \S3 plane merely reflects a
concentration of clusters to the GP, given the {\it stronger}
cluster alignment along the \S3 plane,
it seems a fair (and as we show below, {\it interesting})
exercise to at least consider
the converse supposition.

\section{Discussion}

How the relatively high $Z_{GC}$-distributed, old OC system
was formed has remained a challenging problem.  Among the two most
plausible models, \citet{friel} concludes that old OC creation during 
evolution of the MW disk requires ``fine tuning'' of formation and
destruction processes, whereas in accretion ``one finds a
natural mechanism for open cluster formation'', particularly the high
$|Z_{GC}|$ OCs.  In this context we find several interesting
coincidences regarding the 15 known OC/TC objects with
$d_{\sun}>7$ kpc and $R_{GC}>7$ kpc: (1) Remarkably, 13 of these 15 clusters
are confined to the third and fourth Galactic quadrants, something that
would occur by chance only 4.3\% of the time.  Such a lopsided
distribution is easily accommodated by an accretion origin, but not by a
disk formation model.
(2) Eleven of these 15 clusters are 
among the GASS candidate clusters in Table 1, while one more, Berkeley
22, lies right in the M giant GASS tracing by R03 and would have been
included in our sample had we used a planar distance limit only 20 pc
larger.  Although clearly old (0.7 to 7 Gyr) for OCs, the
Table 1 TC/OC objects are generally poorly studied and have unknown RVs; their
proposed association with one another and with the GASS field star
overdensities must therefore be considered tentative. 
Nevertheless,
their arcing spatial sequence and planar alignment (\S4) is
tantalizingly suggestive of an origin relating to the interaction of a
satellite galaxy with the MW.

\begin{figure}
\epsscale{1.00}
\plotone{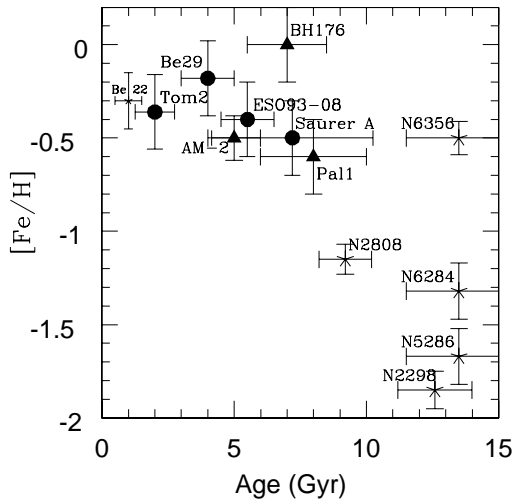}
\caption{AMR of Table 1 clusters plus Be 22 (see \S5).  Symbols as in Fig.\ 2.}
\end{figure}

An alternative view might hold that the arcing spatial sequence of star
clusters and the correlated overdensities of field stars defining GASS
merely represent an outer MW spiral arm.  
However, if a spiral arm,
it is strangely inclined by 17${\arcdeg}$ to the GP.
Additionally, spiral
arms characteristically have {\it young} star clusters, yet no Table 1
cluster is younger than 0.67 Gyr.  
There is also no GASS correlation with the Galactic warp (C03).

On the other hand, 
star clusters in the Fornax (For) and Sgr satellite
galaxies have sizes and luminosities \citep[e.g.,][]{mg03} that span
those of the typical old OC \citep{friel} and Table 1 objects;
that For and Sgr clusters are called ``globular'' seems mainly to
reflect a difference in age, but 
{\it young} clusters of similar
$M_V$ are prevalent \citep{hunter}
in the Magellanic Clouds (MCs).  Under the premise that the Table 1
objects represent the cluster system of a dwarf galaxy, one can use them
to explore the age-metallicity relation (``AMR'') of that putative
system.  Fig.\ 3 shows an AMR typical of that expected for an
independently evolving, ``closed-box'' system with protracted star
formation \citep[compare Fig.\ 3 to the similar AMR of Sgr field stars
and clusters in Fig.\ 18 of][]{layden}.  
Apart from BH 176, the 
Fig.\ 3,
outer TC/OC clusters show a relatively tight AMR (even 
including Be 22), especially compared to that for {\it all} old MW OCs 
\citep[e.g., Fig.\ 8 of][]{friel} and in comparison to the LMC
cluster AMR \citep[e.g., Fig.\ 2a of][]{bica}.  A large metallicity spread among
these clusters mimics the spread among non-cluster GASS stars discussed
by C03.  Together, the various ensemble properties of the GASS clusters
lend further circumstantial support to the ``tidal debris'' explanation
for GASS.  However, while sharing a similar overall AMR to Sgr, GASS appears
to have 
star clusters much younger than in the MW-accreting cluster
system of Sgr, for which the youngest and most metal-rich known cluster
is Terzan 7 (age $8.3\pm1.8$ Gyr, [Fe/H] $=-0.82\pm0.15$; Layden \& Sarajedini 2000).
Clearly such differences in the selective production
and/or destruction of clusters within MW satellites is not a problem
since differences in the distribution of cluster ages are already
observed between the systems in For, Sgr and the MCs, all galaxies with
continuing star formation up to the near present.  But the lower
inclination and apparently smaller orbit of GASS compared to these other
MW satellites suggests consideration of an additional mechanism for the
constant production of new star clusters through the continuous
interaction of a ``GASSeous'' dwarf galaxy with the gaseous disk of the
MW.

We acknowledge funding by NSF grant AST-0307851, NASA/JPL contract
1228235, the David and Lucile Packard Foundation, and the 
Celerity Foundation.

\end{document}